\documentclass[a4paper,12pt]{JHEP3} 

\JHEPspecialurl{http://jhep.sissa.it/JOURNAL/JHEP3.tar.gz}

\usepackage{graphics}
\usepackage{epsfig,multicol,bbm,bm}

\newcommand\fverb{\setbox\pippobox=\hbox\bgroup\verb}
\newcommand\fverbdo{\egroup\medskip\noindent%
			\fbox{\unhbox\pippobox}\ }
\newcommand\fverbit{\egroup\item[\fbox{\unhbox\pippobox}]}
\newbox\pippobox

\title{Star product and the general Leigh-Strassler deformation}

\author{Daniel Bundzik \\
        School of Technology and Society, Malm\"{o} University, \\  \"{O}stra Varvsgatan 11A, S-205 06 Malm\"{o}, Sweden \vspace{0.2cm}
        \\   Department of Theoretical Physics, Lund
        University,\\ S\"{o}lvegatan 14A, S-223 62, Sweden \\
        E-mail: \email{Daniel.Bundzik@ts.mah.se}}

\preprint{\hepth{0608215} \\ LU TP 06-30}	

\abstract{We extend the definition of the star product introduced by Lunin and Maldacena to study marginal deformations of $\mathcal{N}=4$ SYM. The essential difference from the latter is that instead of considering $U(1)\times U(1)$ non-R-symmetry, with charges in a corresponding diagonal matrix, we consider two $\mathbb{Z}_3$-symmetries followed by an $SU(3)$ transformation, with resulting off-diagonal elements. From this procedure we obtain a more general Leigh-Strassler deformation, including cubic terms with the same index, for specific values of the coupling constants. We argue that the conformal property of $\mathcal{N}=4$ SYM is preserved, in both $\beta$- (one-parameter) and $\gamma_{i}$-deformed (three-parameters) theories, since the deformation for each amplitude can be extracted in a prefactor. We also conclude that the obtained amplitudes should follow the iterative structure of MHV amplitudes found by Bern, Dixon and Smirnov.}


\def\tn{\textnormal}
\def\tr{\textnormal{Tr}}
\def\be{\begin{equation}}
\def\ee{\end{equation}}
\def\bea{\begin{eqnarray}}
\def\eea{\end{eqnarray}}

\begin{document} 

\section{Introduction}

The exactly marginal deformations of $\mathcal{N}=4$ supersymmetric Yang-Mills (SYM) preserving $\mathcal{N}=1$ supersymmetry, systematically investigated by Leigh and Strassler in \cite{Leigh:1995ep}, have been studied extensively since the finding, by Lunin and Maldacena in \cite{Lunin:2005jy}, of the supergravity dual of the so-called $\beta$-deformed\footnote{By $\beta$-deformation we mean a one-parameter complex deformation $\beta=\beta_{R}+i\beta_{C}$. With a $\gamma_{i}$-deformed theory we mean a theory containing three complex parameters $\gamma_{1}$, $\gamma_{2}$ and $\gamma_{3}$. In the literature, a $\gamma$-deformed theory sometimes means deformations by the real part of $\beta$ which is called $\beta_{R}$ in the present work.} $\mathcal{N}=4$ SYM theory. Marginal deformations provide an interesting opportunity to study the AdS/CFT-correspondence \cite{Maldacena:1997re} in new supergravity backgrounds.

The perturbative behaviour of the $\beta$-deformed theory shares many features of the undeformed theory \cite{Freedman:2005cg, Penati:2005hp, Mauri:2005pa, Khoze:2005nd}. In \cite{Bern:2005iz} it was found that maximally helicity violating (MHV) planar amplitudes in $\mathcal{N}=4$ SYM have an iterative structure for all $n$-point amplitudes. These results were then transferred to the $\beta$-deformed theory in \cite{Khoze:2005nd} by placing the deformation into the so-called star product. The use of the star product, which was first introduced in this context in \cite{Lunin:2005jy}, to study marginal deformations is especially convenient when calculating amplitudes, since the dependence of the deformation can be isolated into an overall prefactor.

The main purpose of this article is to show that it is possible to obtain the general Leigh-Strassler deformation\footnote{To distinguish from the $\beta$-deformed superpotential we use the word ``general'' when cubic terms of the form $\tr\, \Phi_{i}^{3}$ are present in the Leigh-Strassler deformed theory.}, including cubic terms with all indices equal the same value, from the star product. In section \ref{section-conformal} we discuss the necessary conditions for conformal deformations of $\mathcal{N}=4$ SYM. In Section~\ref{sec3-star-product} we consider two global $\mathbb{Z}_3$-symmetries, in order to solve an eigenvalue system with eigenvectors as a linear combination of the three chiral superfields $\Phi_{i}$. The two systems are related by an element of $SU(3)$ which is also a symmetry of the $\mathcal{N}=4$ SYM Lagrangian written in terms of $\mathcal{N}=1$ superfields. We continue to define the star product for $\mathbb{Z}_3\times \mathbb{Z}_3$-symmetry charges, containing three deformation parameters $\gamma_{i}$. The $\beta$-deformed theory is obtained by putting all parameters equal. In the the diagonal system the star product is easily evaluated. We calculate the superpotential, with ordinary multiplication replaced by the star product, in the $\beta$- and $\gamma_{i}$-deformed theories. The result is  the general Leigh-Strassler deformed superpotential, including the terms of the form $\tr\,\Phi_{i}^{3}$. In section \ref{sec-composite} we compute the starproduct of two chrial superfields which are simple in the $\beta$-deformed case. In appendix \ref{appendix} we present the the results in the $\gamma_{i}$-deformed theory. In section \ref{sec-tree-level} we study the tree-level amplitudes corresponding to terms in the classical Lagrangian. In the $\beta$-deformed theory we find the expected 4-point scalar interaction terms for a Leigh-Strassler deformed theory. However, in the $\gamma_{i}$-deformed case we obtain component terms of the form $\tr\,\phi_{i}^{\dagger}\phi_{i}^{\dagger}\phi_{i}\phi_{j}$, i.e with three identical indices, which are not normally considered in a Leigh-Strassler deformed theory. Their gauge invariance and supersymmetric properties have to be investigated. In Section \ref{sec-phase} we extend the proof in \cite{Khoze:2005nd} which shows
that the phase-dependence of HMV planar tree- and loop-diagrams can be computed from an effective tree-level vertex, determined only by external fields. We conclude that the proof also holds for our present theories. In the final section we compute the one-loop finiteness conditions for conformal marginal deformed $\mathcal{N}=4$ supersymmetric theories with both $\beta$- and $\gamma_{i}$-deformation.
\section{Conformal deformations of $\mathcal{N}=4$ SYM}~\label{section-conformal}
The most general renormalizable $\mathcal{N}=1$ supersymmetric action which is
invariant under a gauge group $G$, can be written as, excluding gauge-fixing and ghost terms,\footnote{We use
the conventions of \cite{Wess&Bagger} such that the generators of
the gauge group satisfy 
$\left[T^{a}_{R},T^{b}_{R}\right]=if^{ab}_{\ \ \ c}T^{c}_{R}$ for
the representation $R$. The adjoint representation $A$ is given by the structure constants such that $\tn{ad}\,
T^{a}_{R}=\left(T^{a}_{A}\right)^{b}_{\ c}=-if^{ab}_{\ \ c}$,
normalized as $\tr\,T^{a}_{A}T^{b}_{A}=-T(A)\delta^{ab}$. }
\be
S= \frac{1}{16T(A)g^2}\int d^{4}xd^{2}\theta \tr
W^{\alpha}W_{\alpha} +\int d^{4}xd^{2}\theta
d^{2}\bar{\theta}\Phi^{\dag}_{A}\left(e^{2gV}\right)^{A}_{\
B}\Phi^{B} +\int d^{4}xd^{2}\theta \mathcal{W} + \tn{h.c}
\label{action}.\ee The chiral superfield $\Phi_A$ and its
conjugate transform under irreducible representations $R$
of $G$. The index $A$ runs over irreducible representations $R_i$ and
the component of each irreducible representation is labeled by $I$,
such that $A=\{i,I\}$ \cite{Parkes:1984dh}. The vector superfield $V^{A}_{\ \
B}=V_{a}\left(T^{a}\right)^{A}_{\ B}$  contains the generators
$T^{a}$, $a=1,\ldots, \tn{dim}\,G$, of the gauge group $G$ defined
by $\left(T^{a}\right)^{A}_{\ B}=\left(T^{ai}\right)^{I}_{\ J}$. The
first term in (\ref{action}) is related to the gauge theory
kinematic Lagrangian containing the gauge field $A^{\mu}$ and a
Majorana spinor, which we call $\lambda_4$. $\mathcal{W}$ is the
superpotential and is given by
\be \mathcal{W}=C_{ABC}\Phi^{A}\Phi^{B}\Phi^{C} \, ,\label{superpotential1} \ee
where $C_{ABC}$ is totally symmetric in $A$, $B$ and $C$ or
equivalent totally symmetric in the pairs $\{i,I\}$, $\{j,J\}$ and
$\{k,K\}$. In the following we will restrict ourselves to
\be
C_{ABC}\equiv
C^{i\,j\,k}_{I\,J\,K}=a^{ijk}b_{IJK}+h^{ijk}d_{IJK} \, ,\label{C-couplings}
\ee where $a^{ijk}$ and $b_{IJK}$ are totally
anti-symmetric and $h^{ijk}$ and $d_{IJK}$ are totally symmetric.

The supercurrent $J_{\alpha\dot{\alpha}}$ of the theory has the anomaly \cite{{Parkes:1984dh},Leigh:1995ep}
\be 
\bar{D}^{\dot{\alpha}}J_{\alpha\dot{\alpha}}=-\frac{1}{3} \left[\frac{\beta_{g}}{g}W^{\beta}W_{\beta}+(d_{s}-3)+\gamma^{i}_{\ j} \left(\Phi_{i} \bar{D}_{\dot{\beta}}\bar{D}^{\dot{\beta}}\Phi^{\dagger j }\right)\right] \, . \label{anomaly-supercurrent}
\ee where $\gamma^{i}_{\ j}$ is the anomalous dimension for $\Phi^{i}$. The anomaly (\ref{anomaly-supercurrent}) is zero in a conformal theory.
At one-loop we have
\be
\beta^{(1)}_{g}=\frac{g^3}{16\pi^2}\left[\sum_{i}T(R_{i})-3C_{2}(G) +\sum_{i,j}T(R_{i})\gamma^{(1)i}_{\ \ \ \;j}\right] \, ,\label{beta-g} 
\ee
and 
\be
\beta^{(1)}_{h_{ijk}}=h_{ijk}\left[(d_{s}-3) -\frac{1}{2}\sum_{i,j}r_{i}\gamma^{(1)i}_{\ \ \ \;j}\right] \, .\label{beta-h}
\ee
The number $r_{i}$ counts the number of chiral fields in each term of the superpotential with the sum $d_s =\sum_{i} r_{i}$. The anomalous dimension is \cite{Jones:1984cx}
\be
\gamma^{(1)i}_{\ \ \ \,j}=C^{ikl}_{IKL}\bar{C}_{jkl}^{JKL}-2g^{2} T(R)\delta^{i}_{j}\delta^{I}_{J} \, .\label{gamma-anomalous}
\ee Vanishing of the one-loop anomalous dimension also implies UV finiteness of $\mathcal{N}=1$ SYM at two-loop level  \cite{Jones:1984cx}.

$\mathcal{N}=4$ supersymmetric Yang-Mills in the $\mathcal{N}=1$ superfield formulation contains three chiral superfields in the
adjoint representation of the $SU(N)$ gauge group and is obtained
by taking $i=1,2,3$ and $I\equiv a=1,\ldots,N^{2}-1$. Thus, if we define $\Phi^{j}\equiv  \Phi^{i}_{a} T^{a}$ the structure constants are $\varepsilon_{IJK}=f_{abc}$, which can be expressed $f_{abc}=-iT(R)^{-1}\tr\,T^{a}\left[T^{b},T^{c}\right]$. The symmetric part $d_{abc}$ vanishes for a real representation.  The $\mathcal{N}=4$ SYM
superpotential becomes
\be \mathcal{W}_{\mathcal{N}=4}=-\frac{ig}{T(R)}\varepsilon^{ijk}\tr\,
\Phi_{i}\left[\Phi_{j},\Phi_{k}\right] \, .\ee  In the Wess-Zumino gauge, the $\mathcal{N}=4$ supersymmetric
Lagrangian can be written in terms of $\mathcal{N}=1$ component fields as

\bea \mathcal{L}&=& \tr \left( \frac{1}{4}F_{\mu\nu}F^{\mu\nu}
-i\lambda_{4}^{\dagger}\bar{\sigma}^{\mu}\mathcal{D}_{\mu}\lambda_{4}- i\lambda_{i}^{\dagger}\bar{\sigma}^{\mu}\mathcal{D}_{\mu}\lambda_{i}-\bar{\mathcal{D}}_{\mu}\phi_{i}^{\dagger}\mathcal{D}^{\mu}\phi_{i} \right.
\nonumber \\ && \left.
-\frac{\sqrt{2}g}{T(A)}\left(\lambda_{4}\left[\phi_{i}^{\dagger},\lambda_{{i}}\right]+
\lambda_{4}^{\dagger}\left[\lambda_{i}^{\dagger},\phi_{i}\right]\right) -
\frac{g}{T(A)}\left(\varepsilon^{ijk}\lambda_{i}\left[\vphantom{\phi_{i}^{\dagger}}\lambda_{j},\phi_{k}\right]+
\varepsilon^{ijk}\lambda_{i}^{\dagger}\left[\lambda_{j}^{\dagger},\phi_{k}^{\dagger}\right]\right) \right.
 \nonumber \\ && \left.
-\frac{g^2}{2\,T(A)^2}\left[\phi_{i}^{\dagger},\phi_{i}\right]\left[\phi_{j}^{\dagger},\phi_{j}\right]
-\frac{2g^2}{\,T(A)^2} \left[\phi_{i}^{\dagger},\phi_{j}^{\dagger}\right]
\left[\vphantom{\phi_{i}^{\dagger}}\phi_{i},\phi_{j}\right] \right) \, .\label{Lagrangian}
 \eea
 Conformal invariance of $\mathcal{N}=4$ SYM follows from (\ref{gamma-anomalous}) where $\gamma^{(1)i}_{\ \ \ \;j}=0$ since $C^{ikl}_{IKL}= g T(R)\epsilon^{ijk}f_{abc}$. This also implies that $\beta^{(1)}_{h_{ijk}}=\beta^{(1)}_{g}=0$.  
 
As we will see, marginal deformations of $\mathcal{N}=4$ SYM which preserve the finiteness condition at one-loop can be obtained by replacing the ordinary multiplication between all fields by an operator called star product. The general form of coupling constants (\ref{C-couplings}) which contains the anti-symmetric part $a^{ijk}$ and the symmetric part $h^{ijk}$ can be written on the form
\be
\mathcal{W}=a^{ijk}\tr\,
\Phi_{i}\left[\Phi_{j},\Phi_{k}\right]+h^{ijk}\tr\,\Phi_{i}\left\{\Phi_{j},\Phi_{k}\right\} \, . \label{superpotential2}
\ee
By choosing the non-zero couplings as  $a^{ijk}=\epsilon^{ijk}\lambda/6$, $h^{123}=\lambda(1-q)/6(1+q)$ and $h^{iii}=h^{\prime}/2$ we obtain the general Leigh-Strassler deformation \cite{Leigh:1995ep,{Bundzik:2005zg}}, also known as the full Leigh-Strassler deformation \cite{Mauri:2006uw},
\be 
\mathcal{W}=h\left(\tr\, \Phi_{1} \Phi_{2} \Phi_{3} -q\tr\, \Phi_{1} \Phi_{3} \Phi_{2}  \right)
+ h^{\prime}\left(
\tr\,\Phi_{1}^{3}+
\tr\,\Phi_{2}^{3}+
\tr\,\Phi_{3}^{3}\right) \label{LS-deformation}  .  
\ee
where $h=2\lambda/(1+q)$.

In the next section we will compute the couplings $h$, $q$ and $ h^{\prime}$ in a star product deformed theory. In section \ref{sec-finiteness} we will evaluate the conditions for the supercurrent in (\ref{anomaly-supercurrent}) to remain anomaly-free.

\section{Deformations from star product}~\label{sec3-star-product}
Introducing the star product has shown to be beneficial in the study of marginal deformations of $\mathcal{N}=4$ SYM \cite{Lunin:2005jy,Khoze:2005nd}. In general, it is not easy to compute the star product of two chiral superfields. To simplify the computation we will in this section solve an eigenvalue system. We continue to define the star product for three deformation parameters. This allows us to compute the superpotential for both $\beta$- and $\gamma_{i}$-deformed theories.

%
\subsection{Eigenvalue system}~\label{Eigenvalue-system}
%
The key idea for this work is to make use of the permutation symmetries of the superpotential to study marginal deformations of $\mathcal{N}=4$ SYM, by introducing a generalized multiplication operator between all fields, which we call ``star product''. When the symmetries permute a set of fields in the original so called $\Phi$-system, it is hard to compute the star product directly. Instead, we rotate the system by an $SU(3)$ transformation into the so called $\Psi$-system in which the symmetries act with diagonal elements. In the $\Psi$-system, the star product can easily be computed.

Let us begin by choosing two symmetries of the superpotential which we denote $S_1$ and $S_2$. In the diagonal $\Psi$-system, the symmetries act as $U(1) \times U(1)$ transformations on the vector $\bm{\Psi}=(\Psi_{1},\Psi_{2},\Psi_{3})$ of chiral superfields accordingly

\be S_{i}: \qquad \bm{\Psi} \qquad \longrightarrow \qquad \mathcal{Q}_{i}\bm{\Psi} \, ,\ee where
\be \mathcal{Q}_{1}=\left(
  \begin{array}{ccc}
    1 & 0 & 0 \\
    0 & e^{-i\varphi_1} & 0 \\
    0 & 0 & e^{i\varphi_1} \\
  \end{array}
\right)\qquad \tn{and}\qquad \mathcal{Q}_{2}=\left(
  \begin{array}{ccc}
    e^{i\varphi_2} & 0 & 0 \\
    0 & e^{-i\varphi_2} & 0 \\
    0 & 0 & 1 \\
  \end{array}
\right)\, \label{diagonal-Q} . \ee
At this stage, $\varphi_1$ and $\varphi_2$ are arbitrary parameters. The superpotential (\ref{superpotential2}) and also the Lagrangian (\ref{Lagrangian}) are invariant under an $SU(3)$ transformation. We introduce the vector $\bm{\Phi}=(\Phi_{1},\Phi_{2},\Phi_{3})$ of chiral superfields such that
\be \bm{\Psi}=T\bm{\Phi} \, ,\qquad \qquad \qquad \qquad T\in SU(3) \, .\ee We now demand that the symmetries $S_1$ and $S_2$ act as permutations of the $\Phi_{i}$'s:
\be S_{i}: \qquad \bm{\Phi} \qquad \longrightarrow \qquad \mathcal{P}_{i}\bm{\Phi} \, ,\ee with 
\be \mathcal{P}_{1}=\left(
  \begin{array}{ccc}
    0 & a_{2} & 0 \\
    0 & 0 & a_{3} \\
    a_{1} & 0 & 0 \\
  \end{array}
\right)\qquad \tn{and}\qquad \mathcal{P}_{2}=\left(
  \begin{array}{ccc}
    0 & 0 & b_{3} \\
    b_{1} & 0 & 0 \\
    0 & b_{2} & 0 \\
  \end{array}
\right)\, \label{permutation-P} , \ee where the parameters $a_{i}$ and $b_{i}$ will be determined below.
The relation between $\mathcal{P}_{i}$ and $\mathcal{Q}_{i}$ is
\be
\mathcal{P}_{i}=T^{-1}\mathcal{Q}_{i}T \, . \label{PQ}
\ee
For the permutation matrices to be elements of SU(3), their elements have to satisfy $i)$ $a_{1}a_{2}a_{3}=1$ and $b_{1}b_{2}b_{3}=1$ and $ii)$ $|a_{i}|^{2}=1$ and $|b_{i}|^{2}=1$. It then follows that $\mathcal{P}_{i}^{3}=1$ which is equivalent to $\mathcal{Q}_{i}^{3}=1$. Thus, the relation (\ref{PQ}) breaks the $U(1) \times U(1)$ symmetry to $\mathbb{Z}_3 \times \mathbb{Z}_3$ with $e^{i\varphi_1}=e^{i\varphi_2}=e^{i2\pi/3}$. For simplicity we define $\alpha=e^{i2\pi/3}$ with inverse $\bar{\alpha}$. The relation $1+\alpha +\bar{\alpha}=0$ will be used repeatedly. As a result, the symmetries $S_1$ and $S_2$ act on the $\Psi_i$'s as
\bea S_1 :&&\qquad\left(\Psi_1 , \Psi_2 , \Psi_3
\right)\qquad \longrightarrow \qquad \left(\Psi_1 ,
\bar{\alpha}\Psi_2 , \alpha \Psi_3 \right)\nonumber \\
S_2:&&\qquad \left(\Psi_1 , \Psi_2 , \Psi_3 \right)\qquad
\longrightarrow \qquad \left(\alpha \Psi_1 ,
\bar{\alpha}\Psi_2 , \Psi_3 \right)\, . \label{S1-S2-action}\eea
These relations will be used when we compute the star product in section \ref{sec-one-parameter}.

The most general solution to (\ref{PQ}) is
\be T=\left(
  \begin{array}{ccc}
    a_{1}t_{1} &  a_{1}a_{2}t_{1}&  t_{1} \\
    \alpha a_{1}t_{2} &\bar{\alpha}a_{1}a_{2}t_{2} &  t_{2}\\
    \bar{\alpha}a_{1}t_{3}  & \alpha a_{1}a_{2}t_{3}  & t_{3}  \\
  \end{array}
\right)\, , \label{T-matrix-solution1}
\ee where $a_{i}$ are the parameters of $\mathcal{P}_{1}$ and $b_{i}=\alpha/a_{i+1}$ in $\mathcal{P}_{2}$. The parameters $t_{1}$, $t_{2}$ and $t_{3}$ have to satisfy $i)$ $3t_{1}t_{2}t_{3}a_{1}^{2}a_{2}(\bar{\alpha}- \alpha )=1$ and $ii)$ $|t_{i}|^2=1/3$ for $T\in SU(3)$. These requirements are fulfilled for  (including the conditions for $\mathcal{P}_{i}\in SU(3)$, see below (\ref{PQ}))
\be
\begin{array}{lll}
 a_{1}= e^{i\theta_{1}}\,, \qquad & a_{2}=e^{i\theta_{2}}=e^{-i(\theta_{1}+\theta_{3})}\,, \qquad &  \qquad   a_{3}= e^{i\theta_{3}} \,, \\
t_{1}=\frac{\displaystyle e^{i\rho_{1}}}{\displaystyle \sqrt{3}}\,,& t_{2}=\frac{\displaystyle e^{i\rho_{2}}}{\displaystyle \sqrt{3}}=\frac{\displaystyle i e^{i(\theta_{3}-\theta_{1}-\rho_{1}-\rho_{3})}}{\displaystyle \sqrt{3}}\,,  &  \qquad t_{3}=\frac{\displaystyle e^{i\rho_{3}}}{\displaystyle \sqrt{3}} \,.
\end{array}
\ee
The transfer matrix becomes
\be T=\frac{1}{\sqrt{3}}\left(
  \begin{array}{ccc}
    e^{i(\theta_{1} +\rho_{1})} &  e^{-i(\theta_{3} -\rho_{1})}  &   e^{i \rho_{1}} \\
     \alpha i e^{i(\theta_{3} -\rho_{1}-\rho_{3})}\qquad & \bar{\alpha}i e^{-i(\theta_{1} +\rho_{1}+\rho_{3})} \qquad &  i e^{i(\theta_{3} -\theta_{1}-\rho_{1}-\rho_{3})} \\
    \bar{\alpha}e^{i(\theta_{1} +\rho_{3})}  & \alpha e^{-i(\theta_{3} -\rho_{3})}  & e^{i \rho_{3}}  \\
  \end{array}
\right)\, . \label{T-matrix-solution2} \ee
If we denote the part of the elements in (\ref{T-matrix-solution2}) by $t_{ij}$ which are dependent of the phases $\theta_{i}$ and $\rho_{i}$, then we can write 
\be \Psi_i =\sum_{j} \alpha^{(i+2)j}t_{ij}\,\Phi_{j}=\sum_{j} \alpha^{(i+2)j}e^{i\rho_{i}}\prod_{\tilde{j}}^{j} e^{i\theta_{\tilde{j}}}\,\Phi_{j} \, .\label{relation-psi-phi}\ee This compact form will be useful in the coming sections. The permutation matrices (\ref{permutation-P}) are 
\be \mathcal{P}_{1}=\left(
  \begin{array}{ccc}
    0 & e^{-i(\theta_{1}+\theta_{3})} & 0 \\
    0 & 0 & e^{i\theta_{3}} \\
    e^{i\theta_{1}} & 0 & 0 \\
  \end{array}
\right)\qquad \tn{and}\qquad \mathcal{P}_{2}=\alpha\left(
  \begin{array}{ccc}
    0 & 0 & e^{-i\theta_{1}} \\
    e^{i(\theta_{1}+\theta_{3})} & 0 & 0 \\
    0 & e^{-i\theta_{3}} & 0 \\
  \end{array}
\right)\, .\label{permutation-P-solution} \ee
The transfer matrix (\ref{T-matrix-solution2}) contains four independent parameters. Two of parameters, $\theta_{1}$ and $\theta_{3}$, are inherited from the permutation symmetry in (\ref{permutation-P-solution}). The remaining two parameters, $\rho_{1}$ and $\rho_{3}$, are coming from the original $\mathcal{N}=4$ SYM $SU(4)$ R-symmetry. It is interesting to note that there does not exist a matrix $T$ which takes $\mathcal{Q}_{i}$ to $\mathcal{P}_{i}$ (see (\ref{PQ})) for continuous parameters. As we will see in the next section, the surviving discrete $\mathbb{Z}_3 \times \mathbb{Z}_3$ symmetry will let us define the star product, which is especially simple to compute in the $\Psi$-system.  Transforming to the $\Phi$-system induces extra cubic terms, of the form $\tr \,\Phi_{i}^{3}$, to the superpotential which correspond to terms in the general Leigh-Strassler deformed theory.

%
\subsection{Definition of star product}
%
We define the star product between two fields $\Psi_i$ and
$\Psi_j$ as, in analogy to \cite{Lunin:2005jy},
\be \Psi_i \star \Psi_j =e^{i \det \widetilde{Q}_{ij}}\,\Psi_i
\cdot \Psi_j \, ,\label{star-product} \ee where $\Psi_i \cdot \Psi_j$ is
an ordinary product and the determinant is defined as
\be \det \widetilde{Q}_{ij} = \left |
  \begin{array}{cc}
   \widetilde{Q}_{i}^{1} & \widetilde{Q}_{i}^{2} \\
    \widetilde{Q}_{j}^{1} & \widetilde{Q}_{j}^{2} \\
  \end{array}
\right |=  \left |
  \begin{array}{cc}
    \tilde{\gamma}_{i}Q_{i}^{1} & \tilde{\gamma}_{i}Q_{i}^{2} \\
   \tilde{\gamma}_{j} Q_{j}^{1} & \tilde{\gamma}_{j}Q_{j}^{2} \\
  \end{array}
\right |= \tilde{\gamma}_{i} \tilde{\gamma}_{j}\det Q_{ij} \, . \label{det}
\ee $(Q_{i}^{1} ,Q_{i}^{2})$ are the ${S}_1 \times {S}_2 $
charges of the fields for the  symmetries ${S}_1$ and ${S}_2$ of
the corresponding superpotential. It will be convenient to rewrite
the three deformation parameters $\tilde{\gamma}_{1}$,
$\tilde{\gamma}_{2}$ and $\tilde{\gamma}_{3}$ as
\be
\gamma_{\,2(i+j)}
=\tilde{\gamma}_{i}\tilde{\gamma}_{j},\qquad 2(i+j)\,\, \mbox{mod}\,\, 3 \, , \label{def-gamma} \ee so that $\gamma_{1}
=\tilde{\gamma}_{2}\tilde{\gamma}_{3}$, $\gamma_{2}
=\tilde{\gamma}_{3}\tilde{\gamma}_{1}$ and $\gamma_{3}
=\widetilde{\gamma}_{1}\widetilde{\gamma}_{2}$. Note that the deformation parameters $\tilde{\gamma}_{i}\tilde{\gamma}_{i}$ also exist. Since they always occur in the combination $\tilde{\gamma}_{i} \tilde{\gamma}_{i}\det Q_{ii}$ where $\det Q_{ii}=0$, the deformations $\tilde{\gamma}_{i}\tilde{\gamma}_{i}$ do not have to be accounted for in calculations.

A deformed multiplication law, such as (\ref{star-product}), is
usually denoted $\star$ and called ``star product". Non-commutative
field theories are often obtained by replacing the ordinary
point-wise product of fields by the Moyal star product, which is
defined by a bidifferential operator over some manifold. In the
present context, the star product may be viewed as generalized
couplings between fields. This is a convenient way to study marginal
deformations of supersymmetric $\mathcal{N}=4$ theories.'

In order to prove that the star product is
associative we have to assume that the elementary fields are defined by (\ref{star-product}) and (\ref{det}) with arbitrary parameters $\tilde{\gamma}_{i}$ and that a composite field of $n$ elementary fields is characterized by the additive property $(\widetilde{Q}_{ij\ldots n}^{1},\widetilde{Q}_{ij\ldots n}^{2})$ where 
\be \widetilde{Q}_{ij\ldots n}^{1,2}= \widetilde{Q}_{i}^{1,2}+\widetilde{Q}_{j}^{1,2}+\cdots +\widetilde{Q}_{n}^{1,2} \ . \label{additive}
\ee 

We can now compute the triple star product
\be \Psi_i \star \Psi_j \star
\Psi_k = e^{i\det \widetilde{Q}_{jk}}\Psi_i \star \left( \Psi_j
\cdot \Psi_k \right) = e^{i (\det \widetilde{Q}_{ij}+\det
\widetilde{Q}_{jk}+\det \widetilde{Q}_{ik})}\Psi_i \cdot \Psi_j
\cdot \Psi_k \, .\label{triple-star} \ee 
The computation of the star product in (\ref{triple-star}) is
associative. The proof is given in appendix \ref{appendixA}. To keep the permutation symmetry of the trace operator also in a star product defined theory we use the short-hand notation
\bea
\tr \,\Psi_i \star \Psi_j & \equiv & \frac{1}{2}\left( \tr \,\Psi_i
\star \Psi_j + \tr \,\Psi_j \star \Psi_i \right)\nonumber \\ &=&\frac{1}{2}e^{\gamma_{\,2(i+j)}\det Q_{ij}}\tr \,\Psi_i \cdot \Psi_j+ \frac{1}{2}e^{-\gamma_{\,2(i+j)}\det Q_{ij}}\tr \,\Psi_j \cdot \Psi_i \, .\eea  In other words, we must symmetrize the trace explicitly before replacing the ordinary multiplication with the star product. The trace for
the triple star product is
\bea \tr \,\Psi_i \star \Psi_j \star
\Psi_k \!\!&=&\!\! \frac {1}{3} \,e^{i(\gamma_{k} \det Q_{ij}
+\gamma_{i} \det Q_{jk}+ \gamma_{j} \det Q_{ik})} 
\nonumber \\  && \times \left[e^{2i\gamma_{i}
\det Q_{kj}} +e^{2i\gamma_{j} \det Q_{ik}}+e^{2i\gamma_{k} \det
Q_{ji}} \right]\tr\,\Psi_i  \Psi_j \Psi_k \label{trace-triple}  \, .
\eea 
When all deformations
parameters are equal we obtain the so-called $\beta$-deformed theory
with $\beta=\gamma_{i}$. If not, we have the three-parameter
$\gamma_{i}$-deformed theory. In section \ref{sec-composite} we will compute the star product $\Phi_{i} \star \Phi_{j}$ of two $\beta$-deformed chiral superfields in the $\Phi$-system. The general results for the $\gamma_{i}$-deformed theory are presented in appendix \ref{appendix}.

%
\subsection{Superpotential in the one-parameter deformed theory}~\label{sec-one-parameter}
%
The $\beta$-deformed theory is obtained by setting all
$\gamma_{i}´$'{s} equal in (\ref{trace-triple}). We use the notation 
$\beta=\gamma_{i}$. From (\ref{S1-S2-action}) we find that the
superfields $\Psi_i$ in the superpotential have charges
\bea
\Psi_1  &:& \qquad\left(Q_{1}^{S_1},Q_{1}^{S_2}\right)=\left(0,1
\right) \nonumber \\
\Psi_2  &:& \qquad\left(Q_{2}^{S_1},Q_{2}^{S_2}\right)=\left(-1,-1
\right) \nonumber \\
\Psi_3  &:& \qquad\left(Q_{3}^{S_1},Q_{3}^{S_2}\right)=\left(1,0
\right) \, . \label{superfield-charges}
\eea 
In the $\Psi$-system it is easy to evaluate the star product. From (\ref{trace-triple}) and (\ref{superfield-charges}) we find
\be \mathcal{W}=\tr \,\Psi_1 \star \Psi_2 \star
\Psi_3 -  \tr \,\Psi_1 \star \Psi_3 \star \Psi_2= e^{i\beta} \,\tr \,\Psi_1 \cdot \Psi_2 \cdot \Psi_3 -e^{-i\beta}\, \tr \,\Psi_1 \cdot
\Psi_3 \cdot \Psi_2 \, .
 \label{trace-psi}\ee 
Since the superpotential transforms as the determinant of the $SU(3)$ $T$-matrix in (\ref{T-matrix-solution2}), we have
\be \mathcal{W} =\tr \,\Psi_1 \star \left[\Psi_2 \stackrel{\star}{,} \Psi_3 \right]=\tr \,\Phi_1 \star \left[\Phi_2 \stackrel{\star}{,} \Phi_3\right] \, .\label{invariant}
\ee
If we use the relation (\ref{relation-psi-phi}) between $\Psi$ and $\Phi$ we find
\be \Psi_i  \Psi_j 
\Psi_k =\sum_{l,\,m,\,n}\alpha^{(i+2)l+(j+2)m + (k+2)n}t_{il}t_{jm}t_{kn}\Phi_{l}
\Phi_{m}\Phi_{n} \, .\ee Performing the trace gives
\be
\tr\,\Psi_i \Psi_j \Psi_k 
=\frac{1}{3}\sum_{l,\,m,\,n}\bar{\alpha}^{l+m+n}\!\left(\alpha^{il+jm + kn}\!+\alpha^{kl+im +
jn}\! +\alpha^{jl+km + in} \right) t_{il}t_{jm}t_{kn}\tr\,\Phi_{l}
\Phi_{m} \Phi_{n} .\ee
To relate to the superpotential
we compute
\be \tr\,\Psi_1  \Psi_2
\Psi_3 =\frac{1}{3}\sum_{l,\,m,\,n}\alpha^{n-l}\left(1+\alpha^{l+m + n}
+\bar{\alpha}^{l+m+n} \right) t_{1l}t_{2m}t_{3n}\tr\, \Phi_{l}
\Phi_{m} \Phi_{n} \, ,\ee which is zero unless $l+m+n=0$ mod $3$.
This implies that the only possible terms are
\bea
\tr\,\Psi_1  \Psi_2 \Psi_3 &=& \frac{i}{\sqrt{3}} \left[ \vphantom{\Phi_{3}^{3}}\bar{\alpha}\tr\, \Phi_{1}
\Phi_{2} \Phi_{3}+\alpha\tr\, \Phi_{1}\Phi_{3}
\Phi_{2} \right. \nonumber \\   &&  \left. +\; e^{i(2\theta_{1}+\theta_{3})}\tr\, \Phi_{1}^{3} +e^{-i(\theta_{1}+2\theta_{3})}\tr\, \Phi_{2}^{3}+ e^{-i(\theta_{1}-\theta_{3})}\tr\, \Phi_{3}^{3} \right] \, . \label{even-part} 
\eea 
In a similar way, we can compute the remaining part of the superpotential (\ref{trace-psi}). The superpotential is invariant under $SU(3)$ so that the phases $\theta_{i}$ can be transformed away by  the field redefinition  
\be 
\Phi_i \longrightarrow e^{i(\theta_{i+1}-\theta_{i})/3}\Phi_i \label{field-redefinition}
\ee

Using (\ref{trace-psi}), (\ref{invariant}), (\ref{even-part}) and (\ref{field-redefinition}) gives the $\beta$-deformed superpotential
\bea 
\tr \,\Phi_1 \star \left[\Phi_2 \stackrel{\star}{,} \Phi_3\right]&=&\frac{-2}{\sqrt{3}}\left[\sin(\beta-\frac{2\pi}{3}) \tr\, \Phi_{1} \Phi_{2} \Phi_{3} +\sin(\beta+\frac{2\pi}{3})\tr\, \Phi_{1} \Phi_{3} \Phi_{2}  \right.
 \nonumber \\ &+&  \left. \sin \beta\left(
\tr\,\Phi_{1}^{3}+
\tr\,\Phi_{2}^{3}+
\tr\,\Phi_{3}^{3}\right)\vphantom{\frac{1}{1}}\right] \label{beta-deformed-superpotential}  .  
\eea 

%
\subsection{Superpotential in the three-parameter deformed theory}
%
In this section we let the three deformation parameters be arbitrary. In a similar way as in the previous section we compute
\be \tr \,\Psi_1 \star \Psi_2 \star
\Psi_3 = \frac{1}{3}\sum_{i,\,j,\,k}\left(e^{ix}\alpha^{k-i}+ e^{iy}\alpha^{j-k}+
e^{iz}\alpha^{i-j}\right)t_{1i}t_{2j}t_{3k}\tr\, \Phi_{i}\, \Phi_{j}\,
\Phi_{k}\, , \ee and
\be \tr \,\Psi_1 \star \Psi_3 \star
\Psi_2 = \frac{1}{3}\sum_{i,\,j,\,k}\left(e^{-ix}\bar{\alpha}^{k-i}+
e^{-iy}\bar{\alpha}^{j-k}+
e^{-iz}\bar{\alpha}^{i-j}\right)t_{1i}t_{3j}t_{2k}\tr\, \Phi_{i}\,
\Phi_{j}\, \Phi_{k} \, ,\ee  where we have introduced
\be x =\gamma_{2}+\gamma_{3}-\gamma_{1}\, ,\qquad y =\gamma_{3}+\gamma_{1}-\gamma_{2} \qquad
\tn{and} \qquad z =\gamma_{1}+\gamma_{2}-\gamma_{3}\, . \label{xyz} \ee
Using (\ref{invariant}) then gives the superpotential
\be \mathcal{W}=\tr \,\Phi_1 \star \left[\Phi_2 \stackrel{\star}{,} \Phi_3\right] =
\frac{2i}{3}\sum_{i,\,j,\,k}P_{i,\,j,\,k}(x,y,z)t_{1i}t_{2j}t_{3k}\tr\, \Phi_{i}\,
\Phi_{j}\, \Phi_{k} \, ,\label{P-superpotential}\ee where 
\be P_{i,\,j,\,k}(x,y,z)=\sin\left(x +(k-i)u\right) +
\sin\left(y+(j-k)u\right)+\sin\left(z+(i-j)u\right)\, . \ee
 with $u=2\pi/3$. Explicitly the terms are
\bea
P_{i,\,i,\,i}(x,y,z)&=&\sin\left(x \right) +
\sin\left(y\right)+\sin\left(z\right)\, , \nonumber\\
 P_{i,\,i+1,\,i+2}(x,y,z)&=&\sin\left(x-u \right) +
\sin\left(y-u\right)+\sin\left(z-u\right) \, ,\nonumber
\\  P_{i,\,i+2,\,i+1}(x,y,z)&=&\sin\left(x+u \right) +
\sin\left(y+u\right)+\sin\left(z+u\right) \, .\label{P-surfaces}\eea The indices are modulus three. All
other terms vanish for any value of $x$, $y$ and $z$, due to the
cyclic property of the trace operator. The $P$-functions\footnote{These functions are not arbitrary named, since the level-set surfaces (\ref{P-surfaces}) belongs to the class of
triply periodic minimal surfaces and are known in the literature as
Schwartz'{s} P-surfaces.} satisfy the
identity
\be
P_{i,\,i,\,i}(x,y,z) +  P_{i,\,i+1,\,i+2}(x,y,z)
+P_{i,\,i+2,\,i+1}(x,y,z) =0 \, .\ee
Finally, after using the field redefinition (\ref{field-redefinition}), the $\gamma_{i}$-deformed superpotential becomes
\bea 
\tr \,\Phi_1 \star \left[\Phi_2 \stackrel{\star}{,} \Phi_3\right]&=& \frac{-2}{\sqrt{3}} \left[  P_{1,2,3}(x,y,z)\tr\, \Phi_{1} \Phi_{2} \Phi_{3} +P_{1,3,2}(x,y,z)\tr\, \Phi_{1} \Phi_{3} \Phi_{2} 
 \nonumber \right. \\   &+& P_{1,1,1}(x,y,z) \left(\tr\,\Phi_{1}^{3} + \tr\,\Phi_{2}^{3} + 
\tr\,\Phi_{3}^{3}\right) ]  \label{gamma-deformed-superpotential}   \, . \eea
The superpotential (\ref{gamma-deformed-superpotential}) is of the form of the general Leigh-Strassler deformation (\ref{LS-deformation}) which can be seen by defining
\be
h =\frac{-2}{\sqrt{3}} P_{1,2,3}(x,y,z) \ , \qquad
q = -\frac{P_{1,3,2}(x,y,z)}{ P_{1,2,3}(x,y,z)}  \ ,  \qquad
h^{\prime} =\frac{-2}{\sqrt{3}} P_{1,1,1}(x,y,z) \ .
\ee

%
\section{Star product of composite chiral superfields}~\label{sec-composite}
%
It is straightforward to compute the star product of two chiral superfields in the $\Phi$-system. These relations are useful when evaluating Feynman diagrams.
To begin, we recall (\ref{relation-psi-phi}) with inverse
\be \Phi_{i} =\sum_{j}\,\bar{\alpha}^{(i+2)j}t_{ji}^{*}\,\Psi_j =\sum_{j}\,\bar{\alpha}^{(i+2)j}e^{-i(\rho_{j}+\sum_{i}\theta_{i})}\,\Psi_j \, .
\ee 
which gives the star product 
\bea \Phi_{i}\star
\Phi_{j}&=&\frac{1}{9}\,\sum_{k,l,m, n}\alpha^{(k+2)(m-i)+(l+2)(n-j)}e^{i\gamma_{2(k+l)}\det Q_{kl}} e^{i(\sum_{\tilde{m}}^{m}\theta_{\tilde{m}}+\sum_{\tilde{n}}^{n}\theta_{\tilde{n}}-\sum_{\tilde{i}}^{i}\theta_{\tilde{i}}-\sum_{\tilde{j}}^{j}\theta_{\tilde{j}})}\,\Phi_{m}
\Phi_{n} \, \nonumber \\ &=&
\frac{1}{9}\,\sum_{k,m, n}\alpha^{(k-1)(m+n-i-j)}\left(1+\alpha^{n-j}e^{i\gamma_{k+2}}+\alpha^{m-i}e^{-i\gamma_{k+2}}\right)\nonumber \\ && \times e^{i(\sum_{\tilde{m}}^{m}\theta_{\tilde{m}}+\sum_{\tilde{n}}^{n}\theta_{\tilde{n}}-\sum_{\tilde{i}}^{i}\theta_{\tilde{i}}-\sum_{\tilde{j}}^{j}\theta_{\tilde{j}})}\,\Phi_{m}
\Phi_{n}\, . \label{Omega-gamma}
 \eea 
In appendix~\ref{appendix} we present the explicit expressions  for the star product in the  $\gamma_{i}$-deformed case. In the $\beta$-deformed case the expression (\ref{Omega-gamma}) is considerable simplified. All terms are zero unless $i+j-m-n =0$ mod $3$ which gives the expressions
\bea \Phi_{i} \star \Phi_{i} &=&\frac{1}{3}\left[\left( 1 +2\cos \beta\right)\Phi_{i} \Phi_{i}+
\left( 1+2\cos(\beta-\frac{2\pi}{3})\right)e^{i(\theta_{1}-\theta_{3}-3\sum_{\tilde{i}}^{i}\theta_{\tilde{i}})}\Phi_{i+1}  \Phi_{i+2} \right. 
\nonumber \\ && \left. +\,\left(1 +2\cos(\beta+\frac{2\pi}{3})\right)e^{i(\theta_{1}-\theta_{3}-3\sum_{\tilde{i}}^{i}\theta_{\tilde{i}})}\Phi_{i+2}  \Phi_{i+1}\right] \, ,\nonumber \\ 
\Phi_{i} \star \Phi_{i+1} &=&\frac{1}{3}\left[\left( 1 +2\cos \beta\right)\Phi_{i} \Phi_{i+1} + \left( 1+2\cos(\beta-\frac{2\pi}{3})\right)\Phi_{i+1}  \Phi_{i} \right. \nonumber \\
&&\left. +\,\left( 1 +2\cos(\beta+\frac{2\pi}{3})\right)e^{-i(\theta_{1}-\theta_{3}-3\sum_{\tilde{i}}^{i+2}\theta_{\tilde{i}})}\Phi_{i+2} \Phi_{i+2} \right] \, , \nonumber \\ 
\Phi_{i+1} \star \Phi_{i} &=&\frac{1}{3}\left[\left( 1 +2\cos \beta\right)\Phi_{i+1} \Phi_{i} + \left( 1 +2\cos(\beta+\frac{2\pi}{3})\right)\Phi_{i}  \Phi_{i+1} \right.\nonumber \\
&&+\left.\,\left( 1
+2\cos(\beta-\frac{2\pi}{3})\right)e^{-i(\theta_{1}-\theta_{3}-3\sum_{\tilde{i}}^{i+2}\theta_{\tilde{i}})}\Phi_{i+2} \Phi_{i+2}\right] \, .
 \eea

%
\section{Tree-level amplitudes from star product}~\label{sec-tree-level}
%
To begin, we replace the ordinary multiplication between all component fields in the Lagrangian (\ref{Lagrangian}) by the star product. From (\ref{superfield-charges}) we find that the component fields have the charges  
\bea
\psi_1 , \lambda_1 &:& \qquad\left(Q_{1}^{S_1},Q_{1}^{S_2}\right)=\left(0,1
\right) \nonumber \\
\psi_2 , \lambda_2 &:& \qquad\left(Q_{2}^{S_1},Q_{2}^{S_2}\right)=\left(-1,-1
\right) \nonumber \\
\psi_3 , \lambda_3 &:& \qquad\left(Q_{3}^{S_1},Q_{3}^{S_2}\right)=\left(1,0
\right) \nonumber \\
A^{\mu}, \lambda_4 &:& \qquad\left(Q_{4}^{S_1},Q_{4}^{S_2}\right)=\left(0,0
\right) \, .
\eea 
The  part 
\be \mathcal{L}_{inv}= -\tr \left( 
\frac{\sqrt{2}g}{T(A)}\left(\lambda_{4}\left[\phi_{i}^{\dagger},\lambda_{{i}}\right]+
\bar{\lambda}_{4}\left[\lambda_{i}^{\dagger},\phi_{i}\right]\right) 
+\frac{g^2}{2\,T(A)^2}\left[\phi_{i}^{\dagger},\phi_{i}\right]\left[\phi_{j}^{\dagger},\phi_{j}\right]
\right) \, ,
 \ee
of the Lagrangian (\ref{Lagrangian}) is unchanged when replacing the normal multiplication with the star product. The reasons are that the gluino $\lambda_{4}$ and its conjugate from the vector multiplet are neutral and that the combinations $\lambda_{i}^{\dagger}\phi_{i}$ and $\phi_{i}^{\dagger}\phi_{i}$, with sum over $i$, are phase-independent.

The terms in the Lagrangian (\ref{Lagrangian}) that are not invariant under the star product are 
\be \mathcal{L}_{\star}= -\frac{g}{T(A)}\tr \left(  
\varepsilon^{ijk}\lambda_{i}\star\left[\lambda_{j}\stackrel{\star}{,}\phi_{k}\right]+
\varepsilon^{ijk}\lambda_{i}\star\left[\lambda_{j}^{\dagger}\stackrel{\star}{,}\phi_{k}^{\dagger}\right]
+\frac{2g}{\,T(A)} \left[\phi_{i}^{\dagger}\stackrel{\star}{,}\phi_{j}^{\dagger}\right]\star
\left[\vphantom{\phi_{i}^{\dagger}}\phi_{i}\stackrel{\star}{,}\phi_{j}\vphantom{\phi_{i}^{\dagger}}\right] \right) \, .\label{star-lagrangian} \ee 
Since the Lagrangian (\ref{Lagrangian}), and naturally (\ref{star-lagrangian}), is invariant under the transformation (\ref{T-matrix-solution2}) we are free to express our fields in the $\psi$-system. From a generalization of the triple star product (\ref{triple-star}) it is easy to evaluate the star product (\ref{star-product}) to express 
\be
\sum_{i,j}\tr\left[\phi_{i}^{\dagger}\stackrel{\star}{,}\phi_{j}^{\dagger} \right]\star 
\left[\vphantom{\phi_{i}^{\dagger}}\phi_{i}\stackrel{\star}{,}\phi_{j}\vphantom{\phi_{i}^{\dagger}}\right]\!\!=2\sum_{i,j,k,l}Q^{ijkl}(\gamma_{1},\gamma_{2},\gamma_{3})\prod_{\tilde{k},\tilde{l},\tilde{i},\tilde{j}}^{k,l,i,j} e^{i(\theta_{\tilde{k}}+\theta_{\tilde{l}}-\theta_{\tilde{i}}-\theta_{\tilde{j}})}
   \tr \, \phi_{i}^{\dagger}\phi_{j}^{\dagger}\phi_{k}\phi_{l} \, , \label{4-scalar-int}
\ee
where we have defined
\be Q^{ijkl} = 
\sum_{m} \left[2\cos \left(2\gamma_{m+2}\! -\!\frac{2\pi n_{1}}{3}\right) - \left(1+\cos 2\gamma_{m+2}\right) \cos \frac{2\pi n_{2}}{3}\right]\alpha^{(m+1)n_{3}} \, , \label{Q} \ee
with
\be n_{1}={i}-{j}-{k}+{l} ,\qquad   n_{2}={i}-{j}+{k}-{l} \qquad \tn{and} \qquad  n_{3}=-{i}-{j}+{k}+{l} \, .\ee
We can see from (\ref{4-scalar-int}) and (\ref{Q}) that interaction terms $\phi_{i}^{\dagger}\phi_{j}^{\dagger}\phi_{k}\phi_{l}$ are allowed for any combination of the indices, in the $\gamma_{i}$-deformed theory. That is, we may have terms with two, three or four indices of the same value. However, in the $\beta$-deformed theory, all terms are proportional to the factor $1+\alpha^{{i}+{j}-{k}-{l}}+\bar{\alpha}^{{i}+{j}-{k}-{l}}$ which is zero unless ${i}+{j}-{k}-{l}~=~0$ mod~$3$. As a consequence, terms with three indices of the same value vanish. In the non-deformed theory, terms with three or four indices of the same value vanish since the interaction is a product of two commutators. Interaction terms with three indices identical are in general not considered in the context of marginal deformations of $\mathcal{N}=4$ SYM. Properties of gauge invariance and supersymmetry have to be investigated.

The four-scalar interaction (\ref{4-scalar-int}) of the $F$-term can be obtained from \be \mathcal{L}_F=\left(\frac{\partial\mathcal{W_{\star}}}{\partial \phi_{i}}\right)^{\dagger}\star \left(\frac{\partial\mathcal{W_{\star}}}{\partial \phi_{i}}\right). \label{F-term}\ee Replacing the star product between the derivatives by an ordinary multiplication, might at first thought give rise to a new theory without terms with three indices of the same value. However, calculations shows that the new couplings are 
\be Q^{ijkl}_{new}=2\sum_{m}\left[\cos (2\gamma_{m+2}-2\pi n_{1}/3)-\cos (2\pi n_{2}/3)\right]\alpha^{(m+1)n_{3}} \label{C-new} \, ,\ee which still contain terms with three identical indices. In obtaining (\ref{C-new}), the trace is not symmetrized since there is an ambiguity how to perform the symmetrization. It might be possible to overcome this ambiguity by evaluating the star product before defining $\Phi^{j}\equiv  \Phi^{i}_{a} T^{a}$ from which it follows that the structure constants $f^{abc}$ are related to the trace operator. This would make (\ref{C-new}) a valid relation. In the present context, the general rule is that all multiplication of fields should be replaced by the star product, as in (\ref{F-term}).

In deriving (\ref{Q}), and also (\ref{C-new}), we have assumed the deformation parameters $\gamma_i$ to be real. To introduce complex variables we can go back to the definition $\gamma_{2(i+j)}=\tilde{\gamma}_{i}\tilde{\gamma}_{j}$, see (\ref{def-gamma}), with $\tilde{\gamma_i}=\tilde{\gamma}_{i}^{R}+i\tilde{\gamma}_{i}^{C}$ where $\tilde{\gamma}_{i}^{R}$ and $\tilde{\gamma}_{i}^{C}$ are real. This leaves us with the deformations
\bea
\tilde{\gamma}_{i}\tilde{\gamma}_{i+1}&=&\tilde{\gamma}_{i}^{R}\tilde{\gamma}_{i+1}^{R}-\tilde{\gamma}_{i}^{C}\tilde{\gamma}_{i+1}^{C}+i\left(\tilde{\gamma}_{i}^{R}\tilde{\gamma}_{i+1}^{C} +\tilde{\gamma}_{i}^{C}\tilde{\gamma}_{i+1}^{R}\right) \equiv  \gamma_{i+2}^{R-}+i\gamma_{i+2}^{C+}\nonumber \\
\tilde{\gamma}_{i}^{*}\tilde{\gamma}_{i+1}&=&\tilde{\gamma}_{i}^{R}\tilde{\gamma}_{i+1}^{R}+\tilde{\gamma}_{i}^{C}\tilde{\gamma}_{i+1}^{C}+i\left(\tilde{\gamma}_{i}^{R}\tilde{\gamma}_{i+1}^{C} -\tilde{\gamma}_{i}^{C}\tilde{\gamma}_{i+1}^{R}\right) \equiv  \gamma_{i+2}^{R+}+i\gamma_{i+2}^{C-}\, , \label{complex-gamma}
\eea
in addition to their complex conjugate. In (\ref{complex-gamma}) there is no obvious way how  to separate the real and imaginary part from our original definition of $\gamma_{i}$ without introducing extra deformations, corresponding to $\tilde{\gamma}_{i}^{*}\tilde{\gamma}_{i+1}$. This complicates the study of the real and complex part of the theory, but might at the same time open up for other interesting possibilities to consider. For complex deformations we find the couplings to be
\bea Q^{ijkl} &=& 
\sum_{m} \left[\cos \left(2\gamma_{m+2}^{R-}-u n_{1}\right)\cosh 2\gamma_{m+2}^{C-} +\cos \left(2\gamma_{i+2}^{R+}-u n_{1}\right)\cosh 2\gamma_{i+2}^{C+}\right. \nonumber \\&-&\left.
\cosh \left(2\gamma_{m+2}^{C+}-iu n_{2}\right) -\cos 2\gamma_{m+2}^{R-}\cos u n_{2}\right]\alpha^{(m+1)n_{3}} \, , \label{complex-couplings}  \eea
where we have used $u=2\pi/3$. If we let $\gamma_{m+2}^{R+}=\gamma_{m+2}^{R-}$ in (\ref{complex-gamma}) and (\ref{complex-couplings}), we obtain the real $\gamma_{i}$-deformed theory with couplings (\ref{Q}), as expected.

To compute the star product of the first term in (\ref{star-lagrangian}), we can make use of the transformation (\ref{T-matrix-solution2}) and the field redefinition (\ref{field-redefinition}) for the component fields $\phi_{i}$ and $\lambda_{i}$. We find 
\bea  \varepsilon^{ijk} \tr\,\lambda_{i}\star\left[\lambda_{j}\stackrel{\star}{,}\phi_{k}\right] &=& \frac{2i}{3}\left(\vphantom{\phi_{i}^{\dagger}} P_{i,i+1,i+2}(x,y,z)\tr\left[  \lambda_{i}\lambda_{i+1}\phi_{i+2}- \lambda_{i}\phi_{i+1}\lambda_{i+2} \right]  \right. \\ &+& \left. P_{i,i+2,i+1}(x,y,z) \tr\left[  \lambda_{i}\lambda_{i+2}\phi_{i+1}- \lambda_{i}\phi_{i+2}\lambda_{i+1} \right]  \right. \nonumber \\ &+& \left.  P_{1,1,1}(x,y,z)\left( \tr  \lambda_{1}\left[\lambda_{1},\phi_{1} \right] + \tr  \lambda_{2}\left[\lambda_{2},\phi_{2} \right] +\tr  \lambda_{3}\left[\lambda_{3},\phi_{3} \right]\right)\vphantom{\phi_{i}^{\dagger}}\right) \, , \nonumber
\eea
where we have used the same notation and definitions as in the equations (\ref{xyz}) and (\ref{P-surfaces}). The conjugate term can be computed in a similar way and equals
\bea  \varepsilon^{ijk} \tr\,\lambda_{i}^{\dagger}\star\left[\lambda_{j}^{\dagger}\stackrel{\star}{,}\phi_{k}^{\dagger}\right] &=& \frac{2i}{3}\left(\vphantom{\phi_{i}^{\dagger}}P_{i,i+2,i+1}(x^{*},y^{*},z^{*})\tr\left[  \lambda_{i}^{\dagger}\lambda_{i+1}^{\dagger}\phi_{i+2}^{\dagger}- \lambda_{i}^{\dagger}\phi_{i+1}^{\dagger}\lambda_{i+2}^{\dagger} \right]  \right.  \nonumber \\ &+& \left. P_{i,i+1,i+2}(x^{*},y^{*},z^{*}) \tr\left[  \lambda_{i}^{\dagger}\lambda_{i+2}^{\dagger}\phi_{i+1}^{\dagger}- \lambda_{i}^{\dagger}\phi_{i+2}^{\dagger}\lambda_{i+1}^{\dagger} \right] \right. \nonumber \\ &+& \left.  P_{1,1,1}(x^{*},y^{*},z^{*})\left( \tr  \lambda_{1}^{\dagger}\left[\lambda_{1}^{\dagger},\phi_{1}^{\dagger} \right]+\tr  \lambda_{2}^{\dagger}\left[\lambda_{2}^{\dagger},\phi_{2}^{\dagger} \right]\right.\right. \nonumber \\ &+& \left.  \left. \tr  \lambda_{3}^{\dagger}\left[\lambda_{3}^{\dagger},\phi_{3}^{\dagger} \right]\right)\vphantom{\phi_{i}^{\dagger}}\right) \, , 
\eea where again the fields have been redefined

%
\section{Phase dependence of amplitudes from star product}~\label{sec-phase}
%
To compute $n$-point loop, or just even tree-level, amplitudes is a tedious work. Organizing the Feynman diagrams by decomposed momentum and helicity, instead of momentum and polarized spin, has shown to dramatically reduce their complexity. These MHV diagrams share an iterative structure for computing higher loops \cite{Bern:2005iz}. Evaluating HMV amplitudes in a star product deformed theory shows the strength of the procedure.

In \cite{Khoze:2005nd} it was shown in a $\beta$-deformed theory not containing terms  $\phi_{i}^{\dagger 2}\phi_{i}^{2}$ that an arbitrary HMV planar tree or loop amplitude has a $\beta$-deformed phase factor which can be read off from a single effective vertex. This vertex is only dependent on the external fields and not on the internal structure. In this section we will show that the results found in \cite{Khoze:2005nd} also hold for our present $\beta$- and $\gamma_{i}$-deformed theories. In doing so, we will briefly extend the proof in \cite{Khoze:2005nd}.

The statement is that the deformation dependence for a general $n$-point HMV planar, tree or loop, amplitude  $\mathcal{A}_{n}(F_{1},\ldots,F_{n})$ is entirely determined by the configuration of the external fields $F_{1},\ldots,F_{n}$, so that
\be
\mathcal{A}_{n}(F_{1},\ldots,F_{n}): \qquad \tr \left( F_{1}\star F_{1} \ldots\star F_{n}\right)= \left[\tn{phase} (\gamma)\right] \tr \left( F_{1} F_{1} \ldots F_{n}\right) \, . \label{effective} \ee
Let us start by considering a general HMV planar tree amplitude. Since an HMV diagram consists of fused vertices of opposite helicity, each propagator is proportional to $F_{I}^{\dagger}\star F_{I}$, with sum over $I$, which is phase independent due to opposite charges. This means that the internal structure is phase independent. A result which is true for both the $\beta$- and the $\gamma_{i}$-deformed theory. Thus, the phase dependence of the amplitude lies entirely in the external fields. 

The argument is the same for planar loop amplitudes. Per definition, a planar diagram has no intersecting lines. Each internal line, between two vertices, is  proportional to $F_{I}^{\dagger}\star F_{I}$, with sum over $I$, which again is independent of the phase. Hence the phase dependence of a planar diagram can be computed from an effective tree-level vertex as in (\ref{effective}), determined only by external fields.

In the $\psi$-system, all planar amplitudes in both the $\beta$- and $\gamma_{i}$-deformed theories are proportional to their $\mathcal{N}=4$ counterparts. Since $\mathcal{N}=4$ SYM is a finite theory, our derived $\beta$- and $\gamma_{i}$-deformed theories should also share the same property of conformal invariance. Since the $\psi$-system is equivalent to the $\phi$-system, through an $SU(3)$ transformation, we can conclude that the Leigh-Strassler deformation obtained from the star product, including diagrams with three indices of the same value, for the specific coupling constants (\ref{P-surfaces}) and (\ref{Q}), are conformal in the planar limit. In the next section we will compute the one-loop finiteness condition. The iterative structure of planar MHV amplitudes in $\mathcal{N}=4$ SYM, studied in \cite{Bern:2005iz}, should also hold for our deformed theories since the phase dependence can be isolated for each amplitude.

%
\section{One-loop finiteness condition}~\label{sec-finiteness}
%
The one-loop finiteness condition is equivalent to the vanishing of the anomalous dimension (\ref{gamma-anomalous}) that was discussed in Section \ref{section-conformal}. If we compare (\ref{C-couplings}) with the superpotential (\ref{P-superpotential}) we find that
\be
C^{ijk}_{abc}=\frac{1}{2}P_{i,j,k}(x,y,z)t_{1i}t_{2j}t_{3k}\left(f^{abc}+d^{abc}\right) \, .
\ee
The antisymmetric property of $f^{abc}$ then gives 
\bea
C^{ikl}_{acd}\bar{C}_{jkl}^{bcd}&=&\frac{1}{4}\sum_{i}\left[\left|P_{i,i+1,i+2}-P_{i,i+2,i+1}\right|^2 f^{acd}f_{bcd} \right. \nonumber \\ &+& \left. \left|P_{i,i+1,i+2}+P_{i,i+2,i+1}\right|^2 d^{acd}d_{bcd} +\left|P_{i,i,i}\right|^{2}d^{acd}d_{bcd} \right] \, .
\eea
Using $f^{acd}f_{bcd}=2N$ and $d^{acd}d_{bcd}=2N-8/N$ and explicitly write the $P$~-~functions in (\ref{P-surfaces}), we find the  one-loop finiteness condition to be
\be g^{2}_{\gamma_{i}} = \frac{3\left|h_{\gamma_{i}}\right|^{2}}{4}\left[3\left|\cos x +\cos y +\cos z\right|^2 +2\left|\sin x +\sin y +\sin z\right|^2 \left(1-\frac{4}{N^2}\right)\right] \, .
\ee
This simplifies to
\be g^{2}_{\beta} = \frac{27\left|h_{\beta}\right|^{2}}{4}\left[3\left|\cos \beta \right|^2 +2\left|\sin \beta \right|^2 \left(1-\frac{4}{N^2}\right)\right] \, .
\ee
in the $\beta$-deformed theory. 
The $\beta$-deformed theory studied in \cite{Khoze:2005nd} showed that a complex deformation of the form $\beta=\beta_{R}+i\beta_{C}$ gives the one-loop finiteness condition $g^2 \propto \left|h\right|^{2}\cosh 2\beta_{C}$ in the large-N limit. Feynman supergraph calculations showed that this planar equivalence with the $\mathcal{N}=4$ SYM theory holds up to four loops.

In the present $\beta$-deformed theory\footnote{Note that here we only have $\beta=\tilde{\beta}\tilde{\beta}$ and $\beta^{*}=\tilde{\beta}^{*}\tilde{\beta}^{*}$. When computing the one-loop conditions, terms as $\tilde{\beta}\tilde{\beta}^{*}$ are not present, so  it is possible to define $\beta=\beta_{R}+i\beta_{C}$ where $\beta_{R}=\beta^{R-}$ and $\beta_{C}=\beta^{C+}$, with notation as in (\ref{complex-gamma}).}, we instead get the planar equivalence
\be
g^{2}_{\beta} \propto \left|h_{\beta}\right|^{2}\left( 2\cosh 2\beta_{C} +\sinh^2 \beta_{C} + \cos^2 \beta_{R}\right) \, ,
\ee
which is dependent on the parameter $\beta_{R}$. It would be interesting to understand the underlying reason for this dependence in a supergraph formalism.
%
\section{Summary and discussion}
%

We have shown that it is possible to obtain the general Leigh-Strassler deformation, including terms of the form $\tr\,\Phi_{i}^{3}$, from the definition (\ref{star-product}) of the star product. The superpotential has been computed for the $\beta$-deformed theory in (\ref{beta-deformed-superpotential}) and for the $\gamma_{i}$-deformed theory in (\ref{gamma-deformed-superpotential}). The analysis was based on two equivalent systems of chiral superfields which we have called the $\Psi$- and the $\Phi$-system, related by an $SU(3)$ transformation. The latter system corresponds to charges in an off-diagonal matrix obtained from an $SU(3)$ transformation of the diagonal $\mathbb{Z}_3 \times \mathbb{Z}_3$-symmetry charges. In the diagonal $\Psi$-system the star product is easily evaluated.

When we computed the tree-level amplitudes corresponding to terms in the classical Lagrangian we found the expected Leigh-Strassler deformed terms for a $\beta$-deformed theory. However, in the $\gamma_{i}$-deformed case, the four-scalar interaction of the $F$-term contained terms of the form  $\tr\,\phi_{i}^{\dagger}\phi_{j}^{\dagger}\phi_{k}\phi_{l}$ for any value of the indices. Terms with three equal indices vanish in the $\beta$-deformed theory, but are present in the $\gamma_{i}$-deformed case. 

We have extended the proof in \cite{Khoze:2005nd} to also cover our present theories. We concluded that for an arbitrary HMV planar tree or loop amplitudes, the phase dependence of the deformation can be computed from an effective tree-level vertex determined only by external fields, and not the internal structure. In the $\psi$-system (component fields) all planar amplitudes in our present theories are proportional to their $\mathcal{N}=4$ counterparts. Since $\mathcal{N}=4$ SYM is a finite theory our present theories should share the same properties. We also concluded that the iterative structure of MHV amplitudes in $\mathcal{N}=4$ SYM, found in \cite{Bern:2005iz}, should also hold for our deformed theories. In section~\ref{sec-finiteness} we computed the one-loop finiteness condition. It would be interesting to find permutation matrices (\ref{permutation-P}) of a more general form to establish a relation between coupling constants and more general conditions for a finite theory.

The supergravity dual to the real $\beta$-deformed theory was  generated in \cite{Lunin:2005jy}, by a combination of T-dualites and a shift (called TsT-transformation) on the isometries of the five-sphere part of $AdS_{5}\times S^{5}$. The complex part of $\beta$ followed from a non-trivial S-duality transformation. In \cite{Frolov:2005dj} for bosons and including fermions in \cite{Alday:2005ww}, it was shown that three consecutive TsT-transformations generate a three-parameter deformation of $AdS_{5}\times S^{5}$. The dual field theory corresponds to a non-supersymmetric three-parameter marginal deformation of $\mathcal{N}=4$ SYM. It would be interesting to understand if the three-parameter supergravity background can be obtained in a similar way, by consecutive TsT-transformations, for our present theories. 

A Lax representation, which implies integrability of strings moving in the Lunin-Maldacena background \cite{Lunin:2005jy}, was also found in \cite{Frolov:2005dj}. In \cite{Berenstein:2004ys} and \cite{Freyhult:2005ws}, it was concluded that the integrability is lost in the planar limit, for complex $\beta$-deformed theories. More general Leigh-Strassler deformed theories, containing $\tr\,\Phi_{i}^{3}$, where consider in \cite{Bundzik:2005zg} to study integrability. It would also be interesting to understand if the present results can be translated to a one-loop dilation operator to win insight in the integrability of marginal deformed $\mathcal{N}=4$ SYM.

\bigskip

\textbf{Acknowledgements}

I would like to thank Anna Tollst\'{e}n for many useful discussions and for reading this manuscript. I would also like to thank Johan Bijnens for discussions and useful inputs.

\appendix 
%
\section{Associativity of the star product}\label{appendixA}
%
In this appendix we will show that 
\be
 \left(\Psi_i \star \Psi_j \right)\star
\Psi_k = \Psi_i \star \left(\Psi_j \star
\Psi_k \right) \ , \label{associative}\ee
which is to say that the star product (\ref{star-product}) is associative.

We begin to use the definition (\ref{additive}) for a composite field of two fields

\be \widetilde{Q}_{ij}^{1}\equiv\widetilde{Q}_{i}^{1}
+\widetilde{Q}_{j}^{1}\ , \qquad \tn{and}\qquad\widetilde{Q}_{ij}^{2}\equiv\widetilde{Q}_{i}^{2}
+\widetilde{Q}_{j}^{2} \ ,
\ee
so that $\Psi_i \cdot \Psi_j$ is characterized by  $(\widetilde{Q}_{ij}^{1},\widetilde{Q}_{ij}^{2})$. The triple star product becomes

\be \Psi_i \star \left(\Psi_j \star
\Psi_k \right)= e^{i\det \widetilde{Q}_{jk}}\Psi_i \star \left( \Psi_j
\cdot \Psi_k \right) = e^{i (\det \widetilde{Q}_{jk}+\det
\widetilde{Q}_{i,jk})}\Psi_i \cdot \Psi_j
\cdot \Psi_k \, ,\label{triple-star1} \ee
where
\bea \det \widetilde{Q}_{i,jk} &\equiv& \left |
  \begin{array}{cc}
   \widetilde{Q}_{i}^{1} & \widetilde{Q}_{i}^{2} \\
    \widetilde{Q}_{jk}^{1} & \widetilde{Q}_{jk}^{2} \\
  \end{array}
\right |=  \left |
  \begin{array}{cc}
    \widetilde{Q}_{i}^{1} & \widetilde{Q}_{i}^{2} \\
   \widetilde{Q}_{j}^{1}
+\widetilde{Q}_{k}^{1} & \widetilde{Q}_{j}^{2}
+\widetilde{Q}_{k}^{2} \\
  \end{array}
\right | \nonumber \\ &=& \widetilde{Q}_{i}^{1}\left(\widetilde{Q}_{j}^{2}
+\widetilde{Q}_{k}^{2}\right) - \widetilde{Q}_{i}^{2}\left( \widetilde{Q}_{j}^{1}
+\widetilde{Q}_{k}^{1}\right)= \widetilde{Q}_{i}^{1}\widetilde{Q}_{j}^{2} - \widetilde{Q}_{i}^{2} \widetilde{Q}_{j}^{1}
+\widetilde{Q}_{i}^{1}\widetilde{Q}_{k}^{2} 
-\widetilde{Q}_{i}^{2}\widetilde{Q}_{k}^{1} \nonumber \\ &=& \det \widetilde{Q}_{ij} +\det \widetilde{Q}_{ik} \, .
\eea
Thus, we have 
\be
\Psi_i \star \left(\Psi_j \star
\Psi_k \right)=  e^{i(\det \widetilde{Q}_{ij}+\det \widetilde{Q}_{jk}+\det 
\widetilde{Q}_{ik})}\Psi_i \cdot \Psi_j
\cdot \Psi_k \, .\label{triple-star2} \ee

To prove associativity we also have to compute
\be \left(\Psi_i \star \Psi_j \right)\star
\Psi_k = e^{i\det \widetilde{Q}_{ij}}\left(\Psi_i \cdot\Psi_j \right)
\star \Psi_k  = e^{i (\det \widetilde{Q}_{jk}+\det
\widetilde{Q}_{ij,k})}\Psi_i \cdot \Psi_j
\cdot \Psi_k \, ,\label{triple-star3} \ee
where
\be \det \widetilde{Q}_{ij,k} \equiv \left |
  \begin{array}{cc}
   \widetilde{Q}_{ij}^{1} & \widetilde{Q}_{ij}^{2} \\
    \widetilde{Q}_{k}^{1} & \widetilde{Q}_{k}^{2} \\
  \end{array}
\right |=  \left |
  \begin{array}{cc}
    \widetilde{Q}_{i}^{1} +\widetilde{Q}_{j}^{1}& \widetilde{Q}_{i}^{2}+\widetilde{Q}_{j}^{2} \\
\widetilde{Q}_{k}^{1} & 
\widetilde{Q}_{k}^{2} \\
  \end{array}
\right | =  \det \widetilde{Q}_{ik} +\det \widetilde{Q}_{jk} \, .
\ee
This means that 
\be
 \left(\Psi_i \star \Psi_j \right)\star
\Psi_k=  e^{i(\det \widetilde{Q}_{ij}+\det \widetilde{Q}_{jk}+\det 
\widetilde{Q}_{ik})}\Psi_i \cdot \Psi_j
\cdot \Psi_k \, .\label{triple-star4} \ee
Comparing (\ref{triple-star2}) and (\ref{triple-star4}) proves the associativity (\ref{associative}) of the star product.

%
\section{Star product in $\gamma_{i}$-deformed theory}\label{appendix}
%
In this appendix we present the results of star product evaluation of two chiral superfields. We us the same notation as in section \ref{sec-composite}. In the $\gamma_{i}$-deformed case we find
\bea
\Phi_{i}&\! \star & \!\Phi_{i}=\frac{1}{9}\sum_{j,k}\left[ \alpha^{(k-1)(i-j)} (1+2\cos \gamma_{k})\prod_{\tilde{i},\tilde{j}}^{i,j}e^{-2i(\theta_{\tilde{i}}-\theta_{\tilde{j}})}  \Phi_{j}\Phi_{j} +\alpha^{(k-1)(i-j+1)} \right.  \\ 
&&\!\!\!\! \left. \times\prod_{\tilde{i},\tilde{j}}^{i,j}e^{-i(2\theta_{\tilde{i}}+\theta_{\tilde{j}})}  \left(\vphantom{\sum}(1+2\cos (\gamma_{k}-u))  \Phi_{j}\Phi_{j+1} +  (1+2\cos (\gamma_{k}+u))  \Phi_{j+1}\Phi_{j} \vphantom{\sum}\right)\right] \, , \nonumber
\eea
\bea
\Phi_{i} &\! \star & \!\Phi_{i+1}=\frac{1}{9}\sum_{j,k}\left[ \alpha^{(k-1)(i-j-1)} (1+2\cos( \gamma_{k}+u))\prod_{\tilde{i},\tilde{j}}^{i,j}e^{i(\theta_{\tilde{i}+2}-2\theta_{\tilde{j}})}   \Phi_{j}\Phi_{j}  \right.\\ 
+&& \!\!\!\!\!\!\left. \alpha^{(k-1)(i-j)} \prod_{\tilde{i},\tilde{j}}^{i,j}e^{i(\theta_{\tilde{i}+2}-\theta_{\tilde{j}+2})} \left(\vphantom{\sum}(1+2\cos \gamma_{k})  \Phi_{j}\Phi_{j+1} +  (1+2\cos (\gamma_{k}-u))  \Phi_{j+1}\Phi_{j} \vphantom{\sum}\right)\right] \, ,\nonumber \\
\Phi_{i+1}\!&\! \star \!& \!\Phi_{i}=\frac{1}{9}\sum_{j,k}\left[ \alpha^{(k-1)(i-j-1)} (1+2\cos( \gamma_{k}-u))\prod_{\tilde{i},\tilde{j}}^{i,j}e^{i(\theta_{\tilde{i}+2}-2\theta_{\tilde{j}})}   \Phi_{j}\Phi_{j}  \right.  \\ 
+&& \!\!\!\!\!\!\left. \alpha^{(k-1)(i-j)}\prod_{\tilde{i},\tilde{j}}^{i,j}e^{i(\theta_{\tilde{i}+2}-\theta_{\tilde{j}+2})}\left(\vphantom{\sum}(1+2\cos (\gamma_{k}+u))  \Phi_{j}\Phi_{j+1} +  (1+2\cos \gamma_{k})  \Phi_{j+1}\Phi_{j} \vphantom{\sum}\right)\right] \, . \nonumber
\eea

 For products involving conjugate superfields we find
\bea
\Phi_{i}\star \Phi_{i}^{\dagger} &=&\frac{1}{9}\sum_{j,k}\left[ \left(3+2 \cos\left(\gamma_{k}-\frac{2\pi}{3}(i-j)\right)\right)  \Phi_{j}\Phi_{j}^{\dagger} \right. \nonumber \\ 
&& \left.+ 2\bar{\alpha}^{k-1} \cos\left(\gamma_{k}-\frac{2\pi}{3}(i-j+1)\right)\prod_{\tilde{j}}^{j}e^{i(\theta_{\tilde{j}}-\theta_{\tilde{j}+1})}  \Phi_{j}\Phi_{j+1}^{\dagger} \nonumber \right.  \\ 
&& \left. + 2\,\alpha^{k-1} \cos\left(\gamma_{k}-\frac{2\pi}{3}(i-j+1)\right)\prod_{\tilde{j}}^{j}e^{-i(\theta_{\tilde{j}}-\theta_{\tilde{j}+1})} \Phi_{j+1}\Phi_{j}^{\dagger}   \right] \, ,
\eea

\bea
\Phi_{i}\star \Phi_{i+1}^{\dagger} &=&\frac{1}{9}\prod_{\tilde{i}}^{i}e^{-i(\theta_{\tilde{i}}-\theta_{\tilde{i}+1})}\sum_{j,k}\left[ 2\alpha^{k-1} \cos\left(\gamma_{k}-\frac{2\pi}{3}(i-j-1)\right)  \Phi_{j}\Phi_{j}^{\dagger} \right. \nonumber \\ 
&& \left.+ \left(3+2 \cos\left(\gamma_{k}-\frac{2\pi}{3}(i-j)\right)\right) \prod_{\tilde{j}}^{j}e^{i(\theta_{\tilde{j}}-\theta_{\tilde{j}+1})}\Phi_{j}\Phi_{j+1}^{\dagger} \nonumber \right.  \\ 
&& \left. + 2\,\bar{\alpha}^{k-1} \cos\left(\gamma_{k}-\frac{2\pi}{3}(i-j)\right)\prod_{\tilde{j}}^{j}e^{-i(\theta_{\tilde{j}}-\theta_{\tilde{j}+1})} \Phi_{j+1}\Phi_{j}^{\dagger}   \right] \, ,
\eea

  \bea
\Phi_{i+1}\star \Phi_{i}^{\dagger} &=&\frac{1}{9}\prod_{\tilde{i}}^{i}e^{i(\theta_{\tilde{i}}-\theta_{\tilde{i}+1})}\sum_{j,k}\left[ 2\bar{\alpha}^{k-1} \cos\left(\gamma_{k}-\frac{2\pi}{3}(i-j-1)\right)  \Phi_{j}\Phi_{j}^{\dagger} \right. \nonumber \\ 
&& \left.+ 2\alpha^{k-1}\cos\left(\gamma_{k}-\frac{2\pi}{3}(i-j)\right)\prod_{\tilde{j}}^{j}e^{i(\theta_{\tilde{j}}-\theta_{\tilde{j}+1})} \Phi_{j}\Phi_{j+1}^{\dagger} \nonumber \right.  \\ 
&& \left.+ \left(3+ 2\, \cos\left(\gamma_{k}-\frac{2\pi}{3}(i-j)\right)\right)\prod_{\tilde{j}}^{j}e^{-i(\theta_{\tilde{j}}-\theta_{\tilde{j}+1})} \Phi_{j+1}\Phi_{j}^{\dagger}   \right] \, ,
\eea

\bea
\Phi_{i}^{\dagger}\star \Phi_{i} &=&\frac{1}{9}\sum_{j,k}\left[\left(3+ 2\, \cos\left(\gamma_{k}+\frac{2\pi}{3}(i-j)\right)\right) \Phi_{j}^{\dagger}\Phi_{j} \right. \nonumber \\ 
&& \left.+ 2\alpha^{k-1}\cos\left(\gamma_{k}+\frac{2\pi}{3}(i-j+1)\right)\prod_{\tilde{j}}^{j}e^{-i(\theta_{\tilde{j}}-\theta_{\tilde{j}+1})} \Phi_{j}^{\dagger}\Phi_{j+1} \nonumber \right.  \\ 
&& \left.+ 2\bar{\alpha}^{k-1} \cos\left(\gamma_{k}+\frac{2\pi}{3}(i-j+1)\right)\prod_{\tilde{j}}^{j}e^{i(\theta_{\tilde{j}}-\theta_{\tilde{j}+1})} \Phi_{j+1}^{\dagger}\Phi_{j}   \right] \, ,
\eea 
 
\bea
\Phi_{i}^{\dagger}\star \Phi_{i+1} &=&\frac{1}{9}\prod_{\tilde{i}}^{i}e^{i(\theta_{\tilde{i}}-\theta_{\tilde{i}+1})}\sum_{j,k}\left[ 2\bar{\alpha}^{k-1} \cos\left(\gamma_{k}+\frac{2\pi}{3}(i-j-1)\right)e^{-i\theta_{3}}  \Phi_{j}^{\dagger}\Phi_{j} \right. \nonumber \\ 
&& \left.+ \left(3+ 2\, \cos\left(\gamma_{k}+\frac{2\pi}{3}(i-j)\right)\right)\prod_{\tilde{j}}^{j}e^{-i(\theta_{\tilde{j}}-\theta_{\tilde{j}+1})} \Phi_{j}^{\dagger}\Phi_{j+1} \nonumber \right.  \\ 
&& \left.+ 2\alpha^{k-1}\cos\left(\gamma_{k}+\frac{2\pi}{3}(i-j)\right)\prod_{\tilde{j}}^{j}e^{i(\theta_{\tilde{j}}-\theta_{\tilde{j}+1})} \Phi_{j+1}^{\dagger}\Phi_{j}   \right] \, ,
\eea 

\bea
\Phi_{i+1}^{\dagger}\star \Phi_{i} &=&\frac{1}{9}\prod_{\tilde{i}}^{i}e^{-i(\theta_{\tilde{i}}-\theta_{\tilde{i}+1})}\sum_{j,k}\left[ 2\alpha^{k-1} \cos\left(\gamma_{k}+\frac{2\pi}{3}(i-j-1)\right)e^{i\theta_{1}}  \Phi_{j}^{\dagger}\Phi_{j} \right. \nonumber \\ 
&& \left.+ 2\bar{\alpha}^{k-1}\cos\left(\gamma_{k}+\frac{2\pi}{3}(i-j)\right)\prod_{\tilde{j}}^{j}e^{-i(\theta_{\tilde{j}}-\theta_{\tilde{j}+1})} \Phi_{j}^{\dagger}\Phi_{j+1} \nonumber \right.  \\ 
&& \left.+ \left(3+ 2\, \cos\left(\gamma_{k}+\frac{2\pi}{3}(i-j)\right)\right)\prod_{\tilde{j}}^{j}e^{i(\theta_{\tilde{j}}-\theta_{\tilde{j}+1})} \Phi_{j+1}^{\dagger}\Phi_{j}   \right] \, .
\eea

\providecommand{\href}[2]{#2}\begingroup\raggedright\endgroup

\end{document}